\documentstyle[epsf,prd,aps,floats]{revtex}
\input{psfig.tex}

\newcommand {\be}{\begin{equation}}
\newcommand {\ee}{\end{equation}}
\newcommand {\bea}{\begin{eqnarray}}
\newcommand {\eea}{\end{eqnarray}}

\begin{document}
\twocolumn[\hsize\textwidth\columnwidth\hsize\csname @twocolumnfalse\endcsname

\title{A Simple Varying-alpha Cosmology}

\author{H\aa vard Bunes Sandvik$^a$, John D.
Barrow$^b$, and Jo\~ao Magueijo$^a$}
\address{{\it a} Theoretical Physics, The Blackett Laboratory,
 Imperial College, Prince Consort Road, London, SW7 2BZ, U.K. \\
{\it b} DAMTP, Centre for Mathematical Sciences, Cambridge
University, Wilberforce Road, Cambridge CB3 0WA, U.K.}

\maketitle

\begin{abstract}
We investigate the cosmological consequences of a simple  theory in
which the electric charge $e$ is allowed to vary. The theory is locally
gauge and Lorentz invariant, and satisfies general covariance. We find
that in this theory the fine structure 'constant',
$\alpha $, remains almost constant in the radiation
era, undergoes small increase in the matter era, but approaches a constant
value when the universe starts accelerating because of the presence of a
positive cosmological constant. This model satisfies geonuclear,
nucleosynthesis, and CMB constraints on time-variation in $\alpha $,
while fitting simultaneously the observed accelerating universe and the
recent high-redshift evidence for small $\alpha $ variations in quasar
spectra.  The model also places specific restrictions on the nature
of the dark matter.
Further tests, involving stellar spectra and
the E\"otv\"os experiment, are proposed.
\end{abstract}

\pacs{PACS Numbers: *** }
] \renewcommand{\thefootnote}{\arabic{footnote}} \setcounter{footnote}{0}

There is renewed interest in cosmological theories where the traditional
constants of Nature can vary in space and time. Theoretical interest is
motivated by string and M-theories in which the true constants exist in more
than $3+1$ dimensions and the effective $(3+1)-$ dimensional constants can
display cosmological variations in time and space. The most observationally
sensitive 'constant' is the electromagnetic fine structure constant, $\alpha
=e^2/\hbar c$, and recent observations motivate the formulation of varying-$%
\alpha $ theories. The new observational many-multiplet technique of Webb et
al, \cite{murphy}, \cite{webb}, exploits the extra sensitivity gained by
studying relativistic transitions to different ground states using
absorption lines in quasar (QSO) spectra at medium redshift. It has provided
the first evidence that the fine structure constant might change with
cosmological time\cite{murphy,webb,webb2}. The trend of these results is
that the value of $\alpha $ was lower in the past, with $\Delta \alpha
/\alpha =-0.72\pm 0.18\times 10^{-5}$ for $z\approx 0.5-3.5.$ Other
investigations have claimed preferred non-zero values of $\Delta \alpha $ $%
<0 $ to best fit the cosmic microwave background (CMB) and Big Bang
Nucleosynthesis (BBN) data at $z\approx 10^3$ and $z\approx 10^{10}$
respectively\cite{avelino,bat}.

A varying $\alpha $ may be closely related to a varying speed of light
(VSL), \cite{moffat93,am,ba}, which is of interest because it appears to
solve the cosmological problems resolved by inflation, together with some
other problems \cite{bm}. VSL theories are usually associated with breaking
of Lorentz invariance (but see \cite{covvsl}). A less radical approach is to
attribute varying $\alpha $ to changes in the fundamental electron charge, $%
e $. Bekenstein \cite{bek2} gives an example of a consistent varying-$e$
theory which preserves local gauge and Lorentz invariance, and is generally
covariant. This is a dilaton theory with coupling to the electromagnetic ``$%
F^2$'' part of the Lagrangian, but not to the other gauge fields. The
approach adopted is only a matter of convenience in the choice of units:
they are observationally indistinguishable \cite{am,barmag98,bek1}.

Another remarkable set of recent observations is of Type Ia supernovae in
distant galaxies. These data have extended the Hubble diagram to redshifts,
$z\ge 1$\cite{super}. They imply an accelerated expansion of the universe.
When combined with CMB data, the supernovae observations favour a flat
universe with approximate matter density, $\Omega _m\approx 0.3$ and vacuum
energy density, $\Omega _\Lambda \approx 0.7$. Studies have attempted to
determine whether quintessential scalar fields could explain both
cosmological dark matter and the recent acceleration of the universe, \cite
{wett,peeb,zlatev,as,rach}.

We will not here attempt to explain the acceleration of the universe.
Instead, we show that by applying a generalisation of Bekenstein's varying-$%
e $ theory in a cosmological setting including the cosmological constant, $%
\Lambda ,$ we are able to explain the magnitude and sense of the observed
change in $\alpha $. The main assumption is that the cold dark matter 
has magnetic fields dominating their electric fields.
The magnetostatic energy then drives changes in $\alpha$ in the matter
dominated epoch, but as the Universe starts to accelerate these changes
become friction dominated and come to a halt. 
This gives a decelerated rate of change in $%
\alpha $, just as the universe starts to accelerate, in accord with both
data sets. The only energy scale we introduce is of the order of Planck
scale, which also makes our model attractive. This model may be seen as a
more conservative alternative to \cite{sn,moffatal}, where a VSL scenario was
proposed which could explain the observed acceleration of the universe and
variations in $\alpha $, as well as their remarkable coincidence in redshift
space.

Bekenstein's original theory takes $c$ and $\hbar $ to be constants and
attributes variations in $\alpha $ to changes in $e$, or the permittivity of
free space. This is done by letting $e$ take on the value of a real scalar
field which varies in space and time $e_0\rightarrow e=e_0\epsilon (x^\mu ),$
where $\epsilon $ is a dimensionless scalar field and $e_0$ is a constant
denoting the present value of $e$. This means some well established
assumptions, like charge conservation, must give way \cite{land}. Still, the
principles of local gauge invariance and causality are maintained, as is the
scale invariance of the $\epsilon $ field.

Since $e$ is the electromagnetic coupling, the $\epsilon $ field couples to
the gauge field as $\epsilon A_\mu $ in the Lagrangian and the gauge
transformation which leaves the action invariant is $\epsilon A_\mu
\rightarrow \epsilon A_\mu +\chi _{,\mu },$ rather than the usual $A_\mu
\rightarrow A_\mu +\chi _{,\mu }.$ The gauge-invariant electromagnetic field
tensor is then $F_{\mu \nu }=\left( (\epsilon A_\nu )_{,\mu }-(\epsilon
A_\mu )_{,\nu }\right)/\epsilon , $ which reduces to the usual form when $%
\epsilon $ is constant. The electromagnetic Lagrangian is still ${\cal L}%
_{em}=-F^{\mu \nu }F_{\mu \nu }/4 $ and the dynamics of the $\epsilon $
field are controlled by the kinetic term ${\cal L}_\epsilon =-\frac 12
\omega (\epsilon _{,\mu }\epsilon ^{,\mu })/{\epsilon ^2}, $ as in \cite
{bek2} (we use a metric with signature -+++). Here the coupling constant $%
\omega$ can be written as $\frac{\hbar c}{l^2}$, where $l$ is the
characteristic length scale of the theory, introduced for dimensional
reasons. This constant length scale gives the scale down to which the
electric field around a point charge is accurately Coulombic. The
corresponding energy scale, $\hbar c/l,$ has to lie above a few tens of $MeV$
to avoid conflict with experiment.

Consider Bekenstein's theory in the cosmological setting suggested by the
recent supernovae results. To simplify calculations, we invoke a
transformation introduced in ref.\cite{joaohaav}. By defining an auxiliary
gauge potential $a_\mu =\epsilon A_\mu ,$ and field tensor $f_{\mu \nu
}=\epsilon F_{\mu \nu }=\partial _\mu a_\nu -\partial _\nu a_\mu ,$ the
covariant derivative takes the usual form,
$D_\mu =\partial _\mu +ie_0a_\mu $.
The dependence on $\epsilon $ in the Lagrangian then occurs only in the
kinetic term for $\epsilon $ and in the $F^2=f^2/\epsilon ^2$ term. 
To simplify further we change
variable: $\epsilon \rightarrow \psi \equiv ln\epsilon .$ The total action
becomes
\begin{equation}
S=\int d^4x\sqrt{-g}\left( {\cal L}_g+{\cal L}_{mat}+{\cal L}_\psi +{\cal L}%
_{em}e^{-2\psi }\right) ,
\end{equation}
where ${\cal L}_\psi =-{\frac \omega 2}\partial _\mu \psi \partial ^\mu \psi
$ and ${\cal L}_{em}=-\frac 14f_{\mu \nu }f^{\mu \nu }$. The gravitational
Lagrangian is the usual ${\cal L}_g=\frac 1{16\pi G}R$, with $R$ the
curvature scalar. Our theory generalises Bekenstein's approach by including
the effects of the varying $\epsilon $ (or $\psi $) field on the
gravitational dynamics of the expanding universe. The scalar field $\psi $
plays a similar role to the dilaton in the low-energy limit of string and
M-theories, with the important difference that it couples only to
electromagnetic energy.
Since the dilaton field couples to {\it all} the matter then the
strong and electroweak charges, as well as particle masses, can
also vary with $x^\mu $. These similarities highlight the deep
connections between effective fundamental theories in higher
dimensions and varying-constant theories, \cite{dims}.

To obtain the cosmological equations we vary the action with respect to the
metric to give the generalised Einstein equations
\begin{equation}
G_{\mu \nu }=8\pi G\left( T_{\mu \nu }^{mat}+T_{\mu \nu }^\psi +T_{\mu \nu
}^{em}e^{-2\psi }\right) ,
\end{equation}
and vary with respect to the $\psi $ field to give the equations of motion
for the field that carries the $\alpha $ variations:
\begin{equation}
\Box \psi =\frac 2\omega e^{-2\psi }{\cal L}_{em}.  \label{boxpsi}
\end{equation}
It is clear that ${\cal L}_{em}$ vanishes for a sea of pure
radiation since then ${\cal L}_{em}=(E^2-B^2)/2=0$. 
This suggests a negligible
change in $e$ in the radiation epoch, a fact confirmed by our numerical
calculations. The only significant contribution to a
variation in $\psi $ comes from nearly pure electrostatic or 
magnetostatic energy.

In the matter epoch changes in $e$ will occur. In order to make quantitative 
predictions we need to know how
non-relativistic matter contributes to the RHS of Eqn.~(\ref{boxpsi}). This
is parametrised by the ratio $\zeta={\cal L}_{em}/\rho $, where
$\rho$ is the energy density, and for baryonic
matter ${\cal L}_{em}\approx E^2/2$. For protons and
neutrons $\zeta_p$ and $\zeta_n$
can be {\it estimated} from the electromagnetic corrections to
the nucleon mass, $0.63$ MeV and $-0.13$ MeV, respectively \cite{zal}.
This correction contains the $E^2/2$ contribution
(always positive), but also terms of the form $j_\mu a^\mu $ (where $j_\mu $
is the quarks' current) and so cannot be used directly.
Hence we take a guiding value $\zeta _p\approx \zeta
_n\sim 10^{-4}$. Furthermore the cosmological value of $\zeta $ (denoted $%
\zeta _m$) has to be weighted by the fraction of matter that is
non-baryonic, a point ignored in the literature \cite{bek2,livio}. Hence, $%
\zeta _m$ depends strongly on the nature of the dark matter, and it could be
that $\zeta _{CDM}\approx -1$ (e.g. superconducting cosmic strings, 
for which ${\cal L}_{em}\approx -B^2/2$), or $|\zeta
_{CDM}|\ll 1$ (neutrinos). BBN predicts an approximate value for the baryon
density of $\Omega _B\approx 0.03$ with a Hubble parameter of $h_0\approx 0.6
$ , implying $\Omega _{CDM}\approx 0.3$. Hence, 
depending on the nature of the dark
matter, $\zeta _m$ can be positive or negative and have a modulus
between $0$ and $\approx 1$.

\begin{figure}[tbp]
\psfig{file=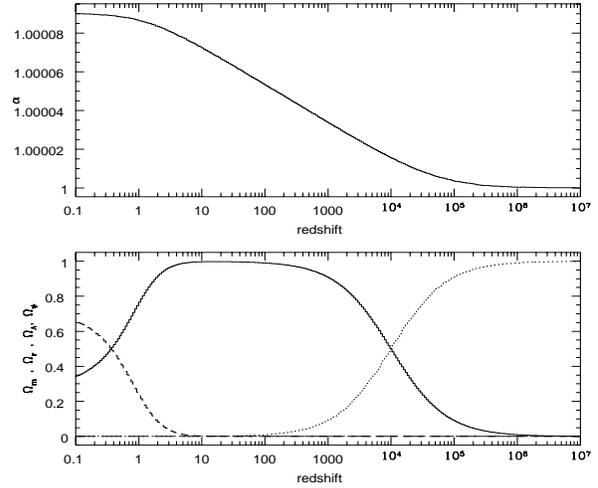,height=7cm,width=8cm}
\caption{Cosmological evolution from radiation domination through
matter domination and into lambda domination with coupling
$\zeta_m/\omega =-0.02\%$. The upper graph plots $\alpha$ as a
function of the redshift $z$.  The lower graph shows the energy
densities of radiation (.....), dust (-----), cosmological
constant (- - -) and the scalar field (combined) as fractions of
the total energy density. The scalar field energy is subdominant
at all times. $\alpha$ increases in the matter era, but approaches
a constant after $\Lambda $ takes over the expansion.}
\label{fig3}
\end{figure}

\begin{figure}[tbp]
\psfig{file=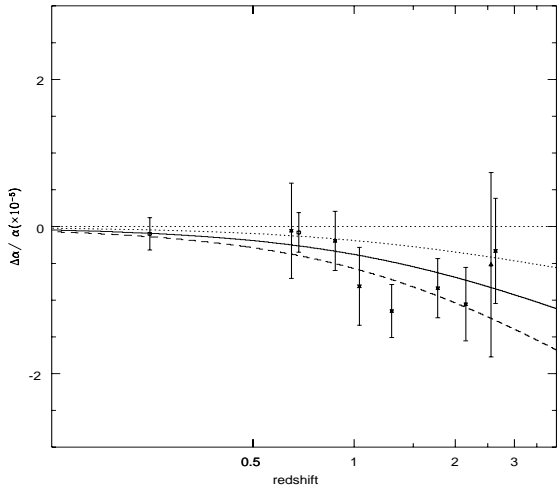,height=7cm,width=8cm}
\caption{The data points are the QSO results for the changing
$\alpha (z)$ reported in refs.\protect\cite{murphy,webb,webb2}.
The solid line is the theoretical prediction for $\alpha (z)$ in
our model with $\zeta _m/\omega
=-0.02\%$. The top (.....) and bottom (- - -) lines correspond to choices $%
\zeta_m/\omega =-0.01\%$ and $\zeta _m/\omega =-0.03\%$
respectively} \label{fig1}
\end{figure}

Assuming a spatially-flat, homogeneous and isotropic Friedmann metric with
expansion scale factor $a(t)$ we obtain the Friedmann equation ($G=c\equiv 1$%
)
\begin{eqnarray}
\left( \frac{\dot{a}}a\right) ^2=\frac{8\pi }3\left( \rho _m\left( 1+\zeta
_me^{-2\psi }\right) +\rho _re^{-2\psi }+\rho _\psi +\rho _\Lambda \right)
\label{fried1}
\end{eqnarray}
where the cosmological vacuum energy $\rho _\Lambda $ is a constant given by
$\Lambda /(8\pi )$, and $\rho _\psi =\frac \omega 2\dot{\psi}^2$. For the
scalar field we get
\begin{equation}
\ddot{\psi}+3H\dot{\psi}=-\frac 2\omega e^{-2\psi }\zeta _m\rho _m
\label{psiddot}
\end{equation}
where $H\equiv \dot{a}/a.$ The conservation equations for the
non-interacting radiation and matter densities, $\rho _r$ and $\rho _m$
respectively, are:
\begin{eqnarray}
\dot{\rho _m}+3H\rho _m &=&0 \\
\dot{\dot{\rho}_r}+4H\rho _r &=&2\dot{\psi}\rho _r.  \label{dotrho}
\end{eqnarray}
This last relation can be written as 
$\dot{\tilde{\rho}_r}+4H\tilde{\rho}_r=0,$ 
with $\tilde{\rho}_r\equiv \rho _re^{-2\psi }\propto a^{-4}$. 

Eqns. (\ref{fried1}-\ref{dotrho}), govern the Friedmann universe with
time-varying $\alpha =\exp (2\psi )e_0^2/\hbar c$. They depend on the choice
of the parameter $\zeta _m/\omega $, which we take to be negative. 
We evolve these equations numerically
from early radiation-domination, through the matter era and into vacuum
domination by $\rho _\Lambda $. 
Fig.~\ref{fig3} shows the evolution of $\alpha $ with redshift in this
model, for $\zeta _m/\omega =-0.02 \%$.
We note that $\psi $ and $\alpha $ remain almost constant during early
radiation domination where baryonic species become relativistic. In the
matter epoch, $\psi $ and $\alpha $ increase slightly towards lower
redshifts, but tend to constant values again once the universe starts
accelerating, and $\Lambda $ dominates - this is due to the friction
term $H\dot \psi$ in Eq.~(\ref{psiddot}). This $\Lambda $ effect reduces
variations in $\alpha $ during the last expansion time of our universe where
the local geonuclear effects of varying $\alpha $ are strongly constrained
by observations, \cite{sh,fujii}, while allowing the cosmological variations
observed by \cite{murphy,webb,webb2} at redshifts, $z\approx 0.5-3.5$, where
the effects of $\Lambda $ on the expansion progressively diminish. In Figure~%
\ref{fig1} we plot the predicted change in $\alpha $
for $-\zeta_m/\omega=0.01, 0.02, 0.03\%$, and the binned QSO data
from refs.\cite{murphy,webb,webb2}.
Given the
uncertainties in $\zeta _m$ discussed above, it is possible to fit
the data with $\omega ={\cal O}(1)$,
so that the theory's length scale is of the order of the Planck
length.

In view of the special $\alpha (z)$ variation produced by the cosmic
acceleration there is agreement with all laboratory, geological and
astrophysical constraints on varying-$\alpha $ deriving from the last
expansion time (cf. \cite{livio,pres,sh,fujii}). Notice also that the
supernovae luminosity data are fitted by our model, since $\psi $ affects
the cosmological expansion very little, and its direct effect upon the
luminosities of astrophysical objects is negligible. Hence, our Hubble
diagram is precisely the same as that of a universe with constant $\alpha $
and $\Omega _m\approx 0.3$ and $\Omega _\Lambda \approx 0.7$. Our model also
meets constraints from BBN, since it occurs deep in the radiation epoch, $%
z\sim 10^9-10^{10},$ when $\alpha $ is predicted to be only $0.007\%$ lower
than today. The standard BBN scenario can withstand variations in $\alpha $
of the order of $1\%$ without contradicting observations (see
\cite{avelino} and references therein).
The value of $\alpha $ at CMB decoupling, $z\approx
1000$ is only $\sim 0.005\%$ lower than today, compatible with
CMB observations, \cite{avelino,bat}, which place an upper bound of a
few percent. However, the variations we predict are close enough to these
limits to hold out the possibility of observational test in the future by
more detailed calculations of the effects on BBN and the CMB, and more
precise data. Low-redshift observations of molecular and atomic transitions
\cite{drink} can provide important information about the value of $\alpha $
close to the redshift where acceleration commences, $z_\Lambda \sim 0.7,$ if
the chemical isotopic evolution uncertainty can be reduced \cite{sel}.

Spatial variations of $\alpha $ are likely to be significant \cite{bt}, and
our model makes firm predictions on how $\alpha $ varies near massive
objects. Linearising eq.~(\ref{boxpsi}) and following the calculation of
\cite{stars}, one finds that relative variations in $\alpha $ are
proportional to the local gravitational potential:
\begin{equation}
{\frac{\Delta \alpha }\alpha }=-{\frac{\zeta _s}{\pi \omega }}{\frac{GM}r}%
\approx 10^{-4}{\frac{\zeta _s}{\zeta _m}}{\frac{GM}r}  \label{star}
\end{equation}
where $M$ is the mass of the compact object, $r$ is its radius, and $\zeta _s
$ is its value of $\zeta $. When $\zeta _m$ and $\zeta _s$ have different
signs, for a cosmologically {\it increasing} $\alpha $, we predict that $%
\alpha $ should {\it decrease} on approach to a massive object. If $|\zeta
_m|\approx \zeta _s$, on extragalactic scales the CMB temperature anisotropy $%
\Delta T/T\sim GM/r$ would lead us to expect large-scale spatial gradients
of order $\Delta \alpha /\alpha \sim 10^{-9}.$ More locally, one would need
an object not larger than some tens of Schwarzschild radii for the effect on
$\alpha (r)$ to be observable with current technology. However with improved
technology, one might find less demanding candidates. An independent low-$z$
test of the effects seen by \cite{murphy,webb} could ultimately be provided
by the detection of a $\Delta \alpha \neq 0$ effect from the fine structure
of stellar spectral lines. The exact relation between the change in $\alpha $
with redshift and in space (near massive objects) is model dependent \cite
{stars}, but eq.(\ref{star}) provides the exact prediction for the simple
varying-$\alpha $ theory considered in this paper. Notice that these
variations do not conflict with the Pound-Rebka-Snider experiments \cite
{will}.

Spatial gradients in $\alpha $ lead to an extra force acting upon matter
coupling to $\psi $ via the $f_{em}^2$ term. Since $\zeta _p\neq
\zeta _n$ this force acts differently on matter with different composition
leading to violations of the weak equivalence principle \cite{zal,ol}.
These are parameterized by the E\"{o}tv\"os parameter, which in our
theory is
\begin{equation}
\eta \equiv {\frac{2|a_1-a_2|}{a_1+a_2}}={\frac{\zeta _E|\zeta _1-\zeta _2|}{%
\omega \pi }}={\frac{\zeta _E|\zeta _1-\zeta _2|}{\pi \zeta _p}}{\frac{\zeta
_p}{\zeta _m}}{\frac{\zeta _m}\omega }
\end{equation}
where $E$ denotes the Earth and 1 and 2 two different test bodies. If $\zeta
_n\approx \zeta _p\approx |\zeta _p-\zeta _n|$ the first factor is ${\cal O}%
(10^{-5})$ for typical substances used in experiments. The third factor,
$\zeta_m/\omega$,  is of the order of $%
- 10^{-4}$. Hence for $|\zeta _m|={\cal O}(0.1)-%
{\cal O}(1)$ we have consistency with the current experimental bound, $\eta
<10^{-13}$ \cite{will}. We note that
the next generation of E\"{o}tv\"os experiments should
be able to detect the variations in $\alpha $ predicted by this theory,
but firmer predictions require better theoretical
calculations of $\zeta $ for neutrons, protons, nuclei and atoms (the
uncertainties of which were discussed above).

In summary, we have shown how a cosmological generalisation of Bekenstein's
theory of a varying $e$ can naturally explain the reported variations in the
fine structure constant whilst satisfying all other observational bounds.
The onset of $\Lambda $ domination is shown to be closely related to the
cosmic epoch when significant changes in $\alpha $ cease to occur. Our
numerical results show that with a natural coupling, and using observational
constraints on the nature of the cold dark matter, $\alpha $ changes
significantly only in the matter dominated epoch. At the onset of $\Lambda $
domination, the expansion accelerates and $\alpha $ rapidly approaches a
constant. This model also places specific restrictions on the nature
of the dark matter. 

Our model complies with geonuclear constraints, like Oklo,
but is consistent with the non-zero variations in $\alpha (z)$ inferred from
observations of quasar absorption lines \cite{murphy,webb,webb2} at $%
z\approx 0.5-3.5$. It is attractive because apart from the (observed)
cosmological constant value, the only free parameter introduced is an energy
scale of the order of the Planck scale. It is also consistent with CMB and
BBN observational constraints. Further tests for this model will be
possible using stellar spectra and the next generation of E\"{o}tv\"os
experiments.

{\bf Acknowledgements} HBS thanks the Research Council of Norway for
financial support. This work was performed on COSMOS, the Origin 2000
supercomputer owned by the UK-CCC. We thank G. Dvali, J. Moffat, M. Murphy,
K. Olive, M. Pospelov, J. Webb and M. Zaldarriaga for discussions.

\end{document}